\documentclass[pra,twocolumn,preprintnumbers,amsmath,amssymb,showkeys]{revtex4}
\usepackage[english]{babel}
\usepackage[utf8]{inputenc}
\usepackage[T1]{fontenc}
\usepackage{lmodern}
\usepackage{textcomp,gensymb}
\usepackage{amsmath}
\usepackage{amssymb}
\usepackage[a4paper]{geometry}
\usepackage{pgf, tikz}
\usepackage{mathtools}
\usepackage{tikz}
\usepackage{graphicx}
\usepackage{dcolumn}
\usepackage{bm}
\usepackage{hyperref}
\usepackage{mathtools}
\usepackage[normalem]{ulem}
\usepackage[super]{nth}
\usepackage{float}

\DeclareMathOperator{\Tr}{Tr}

\DeclareMathOperator{\diag}{diag}

\newcommand{\defeq}{\overset{\makebox[0pt]{\mbox{\normalfont\scriptsize def}}}{=}}

\begin{document}

\title{Landauer-B\"uttiker equation for bosonic carriers}
\author{Andrey~R.~Kolovsky$^{1,2}$}
\author{Zakari Denis$^{3}$}
\author{Sandro~Wimberger$^{4,5}$ }
\affiliation{$^1$Kirensky Institute of Physics, 660036 Krasnoyarsk, Russia}
\affiliation{$^2$Siberian Federal University, 660041 Krasnoyarsk, Russia}
\affiliation{$^3$ Universit\'e Paris Diderot, Sorbonne Paris Cit\'e, 75013 Paris, France}
\affiliation{$^4$ Dipartimento di Scienze Matematiche, Fisiche e Informatiche, Campus Universitario, Parco Area delle Scienze n. 7/a, 
                         Universit\`a  di Parma, 43124 Parma, Italy}                         
\affiliation{$^5$ INFN, Sezione di Milano Bicocca, Gruppo Collegato di Parma, 43124 Parma, Italy}
\date{\today}
\begin{abstract}
We study the current of Bose particles between two reservoirs connected by a one-dimensional channel. We analyze the problem from first principles by considering a microscopic model of conductivity in the noninteracting limit. Equations for the transient and the stationary current are derived analytically. The asymptotic current has a form similar to the Landauer-B\"uttiker equation for electronic current in mesoscopic devices.
\end{abstract}

\pacs{
		03.75.Gg, 
		05.60.-k 
	}

\keywords{Bose-Einstein condensates, ultracold atoms, quantum transport, open quantum systems}

\maketitle

\section{Introduction}
Recent progress in  cold-atom physics witnesses the emergency of the new  field atomtronics \cite{Zoller2004, Seam07, Ott}, which deals with atom-based setups which are similar to crystal-based electronic devices. In particular, the recent series of experiments at ETH \cite{Zurich1,Lebr18} analyzes a current of fermionic atoms between two atomic reservoirs connected by a point contact \cite{Zurich1}, or even by quasi one-dimentional channels with periodic structure \cite{Lebr18}. To explain their experimental results in the case of weakly interacting atoms the authors of \cite{Zurich1,Lebr18} appeal to the Landauer-B\"uttiker (LB) theory \cite{Landau-B}, which has been proved to be very successful in describing ballistic transport of electrons in mesoscopic solid-state devices. 

According to the LB approach \cite{Landau-B} the electron current $j$ is given by the equation 
\begin{equation}
\label{eq:1}
j=\sum_n \int f(E) |t_n(E)|^2 {\rm d} E \,.
\end{equation}
Here $n$ labels the transmission channels, $t_n(E)$ is the transmission amplitude at the energy $E$ of each channel, and the function $f(E)$ is determined by the chemical potentials of the left and right reservoir through the Fermi-Dirac distribution $f_{FD}(E)$ as $f(E)\sim f_{FD}(E-\mu_L)-f_{FD}(E-\mu_R)$. It was demonstrated in the ETH experiments that Eq.~(\ref{eq:1}) describes reasonably well the current of weakly interacting fermionic atoms in engineered optical potentials.  A unique property of atomtronic devices is that they may use Fermi as well as Bose atoms. This rises the question about an analogue of the LB equation for bosonic transport \cite{LB2012}. To obtain such an analogue for particle transport is the main goal of this paper.

\section{Microscopic model}
\begin{figure}[bt]
\includegraphics[width=0.9\linewidth,clip]{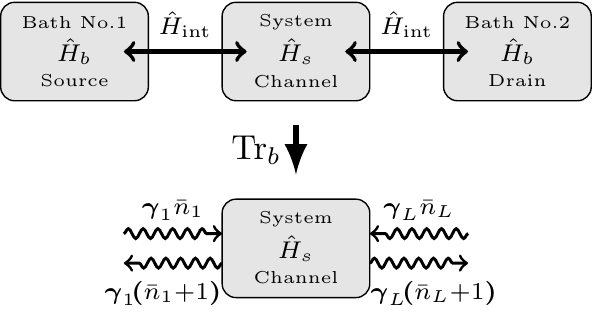}
\caption{Schematic representation of the microscopic model (upper panel) from which the effective dissipative Bose-Hubbard chain description is derived with gain and loss at sites 1 and L (bottom panel).}
\label{fig:1}
\end{figure}

In what follows we illustrate the derivation of the bosonic LB equation by considering a simple microscopic model that is a modification of the model introduced in Ref.~\cite{108}. It consists of two reservoirs of Bose particles connected by the transport channel, see Fig.~\ref{fig:1}(upper panel).  The reservoirs are described by the Bose-Hubbard Hamiltonians $\widehat{H}_b$,
\begin{equation}
\label{eq:2}
\widehat{H}_b=\frac{W}{2}\sum_{m=1}^M\hat{n}_m(\hat{n}_m-1) - \frac{J}{2}\sum_{m=1}^{M-1}\left( \hat{b}^\dagger_{m+1}\hat{b}_{m} + h.c.\right) \;,
\end{equation}
where $W\ne0$ is the interaction constant and we implicitly assume the thermodynamic limit with a fixed filling factor $\bar{n}=N/M$. The reservoirs are connected to the transport channel  described by the Hamiltonian $\widehat{H}_s$,
\begin{align}
\label{eq:3}
\widehat{H}_s=&-\frac{J}{2} \sum_{l=1}^{L-1} \left( \hat{a}^\dagger_{l+1}\hat{a}_{l} + h.c. \right)\nonumber\\
&+ \sum_{l=1}^{L} \left( E_l \hat{n}_l   + \frac{U}{2} \hat{n}_l ( \hat{n}_l - 1 ) \right),
\end{align}
where the on-site energies $E_l$ correspond to a scattering potential.  Finally, the channel is coupled to the left reservoir by the Hamiltonian $\widehat{H}_{int}$
\begin{equation}
\label{eq:4}
\widehat{H}_{int}=\varepsilon\left(\hat{b}_M \hat{a}^\dagger_1 + h.c.\right)   \;,
\end{equation}
where the hopping matrix element $\varepsilon$ plays the role of the coupling constant. The coupling Hamiltonian to the right reservoir (which is  described by its own Bose-Hubbard Hamiltonian) has the form \smash{$\widehat{H}_{int}=\varepsilon(\hat{a}_L \hat{b}^\dagger_1 + h.c.)$}.

To avoid a possible confusion we stress that, in spite of using different notations for the creation and annihilation operators for Bose particles in the transport channel and reservoirs, the total Hamiltonian describes a system of indistinguishable particles. The main assumption of the model is that the interaction constants $W$ and $U$ in the Hamiltonians (\ref{eq:2}) and (\ref{eq:3}) can be varied independently, where we will mainly focus on the case $U=0$ but $W\ne0$. The advantage of using the interacting Bose-Hubbard model \eqref{eq:2} as a bath is that it is generally a quantum chaotic system with universal properties of the energy spectrum and eigenstates \cite{66, Sto2009}. These properties help us to justify the master equation that will be derived below. The advantage of considering a non-interacting channel is that the obtained master equation can be solved analytically without any further approximations.
%

\section{Master equation}
The first step is to derive the master equation for the reduced density matrix describing the carriers in the transport channel,
${\rho}_s(t)={\rm Tr}_b[{\rho}_{tot}(t)]$,
where ${\rm Tr}_b[\ldots]$ denotes the partial trace over the bath variables. A master equation in Lindblad form can be obtained for ${\rho}_s(t)$ within the so-called Born-Markov approximation \cite{Breuer2002}. This approximation assumes that (i) the total density matrix factorizes into the tensor product of the reduced density matrices for all times, 
${\rho}_{tot}(t)={\rho}_b(t)\otimes {\rho}_s(t)$,
%
(here for simplicity we temporally discuss the case of a single bath) and that (ii) bath degrees of freedom are $\delta$-correlated, i.e., 
${\rm Tr}_b [\hat{b}(t)\hat{b}^\dagger(t') {\rho}_b(t=0)]  \sim \delta \left( t-t' \right)$.
We mention, in passing, that the former approximation is actually never satisfied because the coupling Hamiltonian entangles the particles in the channel and the bath. However, one can justify a weaker approximation \cite{25},
${\rm Tr}_b[ F(\hat{b},\hat{b}^\dagger) {\rho}_{tot}(t)] = {\rm Tr}_b[ F(\hat{b},\hat{b}^\dagger)  {\rho}_b(t)] {\rho}_s(t)$,
%
which is sufficient to eliminate the bath ($F(.,.)$ is an arbitrary operator-valued function of the bath variables). The required property behind the previous equation is a quantum counterpart of the mixing property of classically chaotic systems and the interacting Bose-Hubbard model possesses this property \cite{66}.

Using the explicit form of the coupling Hamiltonian (\ref{eq:4}), the result is a Lindblad master equation for the reduced density matrix of the channel, see, e.g., \cite{Ivanov2013}. This provides an effective model for the reduced dynamics in the channel, as sketched in Fig.~\ref{fig:1} (bottom panel). The channel is now a dissipative Bose-Hubbard chain subject to single-particle loss processes at rates $\gamma_{1,L}$ as well as incoherent pumping at the same rates around the populations $\bar{n}_{1,L}$ at both ends, labelled $1$ and $L$, respectively.  Explicitly we have 
\begin{align}
\partial_t\hat{\rho}_s = &-i\big[\widehat{H}_s,\hat{\rho}_s\big]  \nonumber\\
&+\sum_{\mathclap{j=1,L}}\gamma_j\big(\bar{n}_j\mathcal{D}[\hat{a}_j^\dagger]\hat{\rho}_s 
+ (\bar{n}_j+1)\mathcal{D}[\hat{a}_j^{\mathstrut}]\hat{\rho}_s\big)
\label{eq:5}
\end{align}
where the dissipator is defined as $\mathcal{D}[\hat{a}]\hat{\rho}_s = \hat{a}\hat{\rho}_s\hat{a}^\dagger - 1/2\lbrace\hat{a}^\dagger\hat{a},\hat{\rho}_s \rbrace$.
%
%
In terms of the microscopic model of Eqs. (\ref{eq:2}-\ref{eq:4}), the parameters ${\bar n}_1$ and $\bar{n}_L$ are given by the filling factor of the Bose-Hubbard reservoirs and the parameters $\gamma_{1,L}$ are proportional to the square of the lead-channel coupling constant, $\gamma_{1,L} \sim \varepsilon^2$. Notice that, besides the particle exchange between the system and the bath, the relaxation terms in Eq.~\eqref{eq:5} are responsible for decoherence processes as well. The latter is a consequence of the mentioned system-bath entanglement. 

With respect to the microscopic model Eqs. (\ref{eq:2}-\ref{eq:4}), the validity of the master  Eq. \eqref{eq:5} was tested in ref.~\cite{108}. It was found that it is well justified in the high-temperature limit, where the overwhelming majority of bath states are chaotic. However, it may give wrong results in the low-temperature limit, where most of the populated bath states (including the ground state) are regular states. The high-temperature limit (which we shall assume from now on) implies that all Bloch states for a particle in the bath are equally populated. Thus a particle coming into the channel may have an arbitrary quasimomentum. 
Let us also mention that the master Eq.~(\ref{eq:5}) is used as the starting point in many papers on conductivity with bosonic  and fermionic carriers \cite{OtherPapers, Kordas2015,Denis2018}, where in the latter case the bosonic annihilation and creation operator should be substituted by fermionic ones.

\section{Single-particle density matrix}
%
The reduced density matrix ${\rho}_s(t)$ entering the master Eq. (\ref{eq:5}) carries the whole information about the atoms in the transport channel but has a huge dimension which grows exponentially with $L$. Fortunately, for many purposes, it suffices to know only the single-particle density matrix (SPDM), which has the fixed dimension $L\times L$ and is defined as
\begin{equation}
	\sigma_{lm}(t) = \Tr[\hat{\rho}_s(t)\hat{a}_l^\dagger\hat{a}_m^{\mathstrut}] \equiv \langle \hat{a}^\dagger_{l}(t)\hat{a}^{\mathstrut}_m(t) \rangle \,.
	\label{eq:6}
\end{equation}
The SPDM is easily shown to satisfy the following set of equations of motion 
	\begin{align}
	\partial_t\sigma_{lm} &= i(E_l-E_m)\sigma_{lm}\nonumber\\
	&+i\frac{J}{2}(\sigma_{l,m+1}+\sigma_{l,m-1}-\sigma_{l+1,m}-\sigma_{l-1,m})\nonumber\\
	&-i U(\langle \hat{a}_l^\dagger\hat{a}_m^{\mathstrut}\hat{a}_m^\dagger\hat{a}_m^{\mathstrut} \rangle - \langle \hat{a}_l^\dagger\hat{a}_l^{\mathstrut}\hat{a}_l^\dagger\hat{a}_m^{\mathstrut} \rangle)\nonumber\\
	&-\sum_{\mathclap{j=1,L}}\frac{\gamma_j}{2}(\delta_{l,j}+\delta_{m,j})(\sigma_{lm}-\delta_{l,m}\bar{n}_j) \,.
	\label{eq:7}
	\end{align}
These equations can be put into a closed form using an appropriate truncation scheme such as mean field or the beyond-mean-field Bogoliubov back-reaction method \cite{BBR,Kordas2015}. Fortunately, in the case of non-interacting bosons ($U=0$), considered throughout this paper, the equations are already in closed form. This holds for any master equation with Hamiltonian terms and Lindblad operators at most quadratic on the creation and annihilation  operators \cite{Breuer2002}.

\subsection{Transient dynamics of the SPDM}
\begin{figure}[tb]
\includegraphics[width=1.06\linewidth,clip]{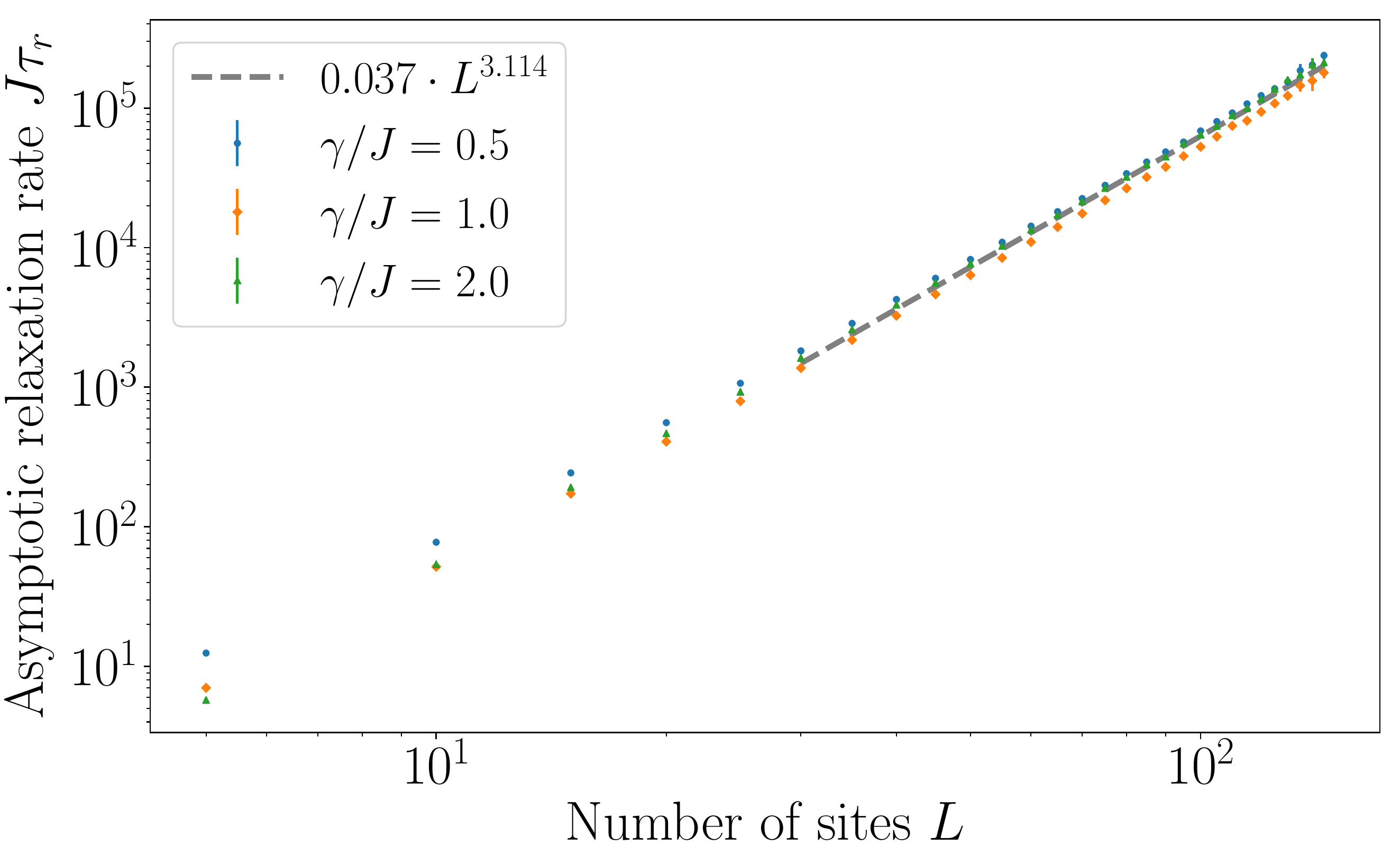}
\caption{\label{fig:2}%
 Asymptotic relaxation time  $\tau_{r}$ vs. $L$ for $\bar{n}_1 = 0.02$, $\bar{n}_L = 0$, and various ratios $\gamma / J$ as shown in the legend. The dashed line shows a fit approximately cubic in $L$.}
\end{figure}

We are interested in the current across the chain. 
By building a vector $\vec{\sigma}$ out of the SPDM's matrix elements, Eq. (\ref{eq:7}) can be rewritten in the form
	\begin{equation}
	\partial_t\vec{\sigma}(t) = \mathbf{A}\vec{\sigma}(t)+\vec{b} \,,
	\label{eq:8}
	\end{equation}
with $\mathbf{A}$ a diagonally dominant complex symmetric matrix. The general solution of such a system is
	\begin{equation}
	\vec{\sigma}(t) = e^{\mathbf{A}t}\vec{\sigma}(0)+{\mathbf{A}}^{-1}(\mathbf{1}-e^{\mathbf{A} t})\vec{b} \,,
	\label{eq:9}
	\end{equation}
which can be explicitly obtained for a few-site chain, for example the two-site system, see appendix \ref{app1} for a detailed discussion of this case. 
However, explicit expressions get more involved for larger systems, for which numerical diagonalization can be efficiently used. This is the reason why we will now study the limit of the stationary current. Indeed, it follows from Eq.~(\ref{eq:9}) that, independent of the initial condition, the SPDM $\sigma_{l,m}(t)$ relaxes to some stationary matrix  $\bar{\sigma}_{l,m}$. Relaxation towards this steady state occurs on a time scale of the order of the asymptotic relaxation time, defined as $\tau_{r} = \inf_\ell(\lbrace\lvert\lambda_\ell\rvert\rbrace)^{-1}$, where $\lbrace\lambda_\ell\rbrace$ denotes the set of eigenvalues of the matrix $\mathbf{A}$ of Eq.~(\ref{eq:8}). We find that the relaxation time $\tau_{r}$ increases with the chain length $L$. This rises the question of its asymptotic scaling. Fig.~\ref{fig:2} shows that 
$\tau_{r}$ scales with the number of sites $L$ as an approximate cubic power law, as predicted in ref. \cite{Zni2015} for boundary dissipation.

\begin{figure}[tb]
\includegraphics[width=\linewidth]{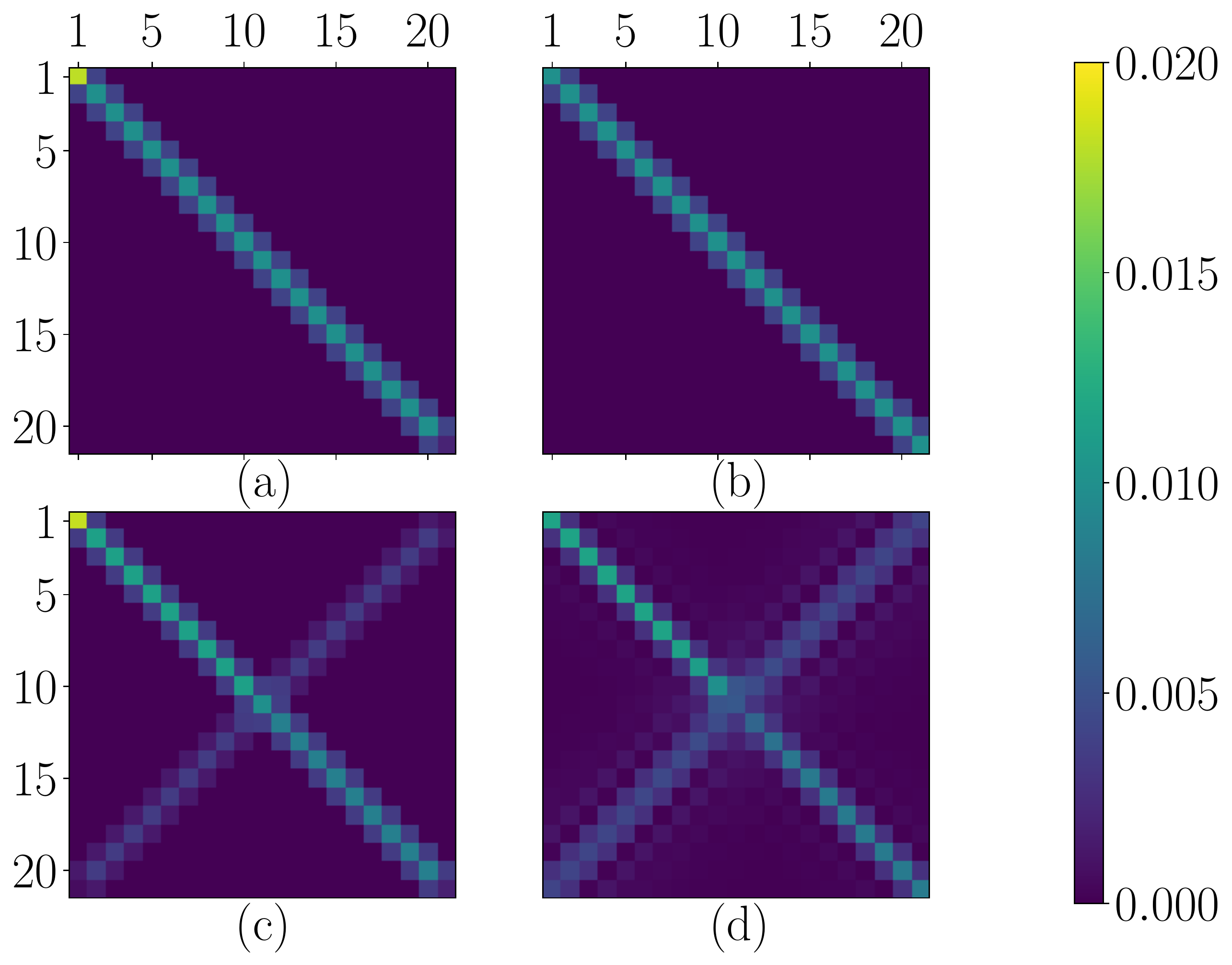} \ \
  \caption{\label{fig:3}%
Absolute values of the steady state SPDM for $L = 21$, $\bar{n}_1=0.02$, $\bar{n}_L=0$, $\gamma /J  = 2$ and either $F/J = 0$ (upper panels) or $F/J = 0.5$ (bottom panels) for the $V_1$ scattering potential. Exact numerical results (left) are to be compared to the LB ones (right), computed from Eq.~\refeq{eq:16}.}
\end{figure}

To analyze the regime of stationary currents, we first consider the case of identical on-site energies $E_l$, where the transmission probability is unity. In this case the SPDM relaxes to a tridiagonal matrix  $\bar{\sigma}_{l,m}$ shown in Fig.~\ref{fig:3}(a). This matrix is uniquely characterized by four quantities: the value of its diagonal elements $A$ (except for the first and last sites), the value of its off-diagonal elements $\pm iB$, and $\bar{\sigma}_{1,1}=C$ and  $\bar{\sigma}_{L,L}=D$. To simplify the equations we shall restrict ourselves to the parameter region $\gamma_1=\gamma_L\equiv \gamma$. Setting the left-hand-side of Eq.~\eqref{eq:7} to zero for $U=0$, we obtain the following system of algebraic equations for the unknown matrix parameters
\begin{eqnarray}
\gamma C + JB = \gamma \bar{n}_1 \;, \label{eq:10}\\
\gamma B - JC +JA = 0 \;, \label{eq:11}\\
\gamma B -JA +JD = 0 \;, \label{eq:12}\\
\gamma D -JB = \gamma \bar{n}_L \;, \label{eq:13}
\end{eqnarray} 
which gives  
\begin{eqnarray}
\label{eq:14}
A=\frac{C+D}{2}=\frac{\bar{n}_1+\bar{n}_L}{2} \;,
\end{eqnarray} 
and 
\begin{eqnarray}
\label{eq:15}
B= \frac{J\gamma}{\gamma^2+J^2}\frac{\bar{n}_1-\bar{n}_L}{2} \;.
\end{eqnarray} 
Keeping in mind the matrix of the current operator, $j_{j,m}=j_0(\delta_{l,m+1}-\delta_{l+1,m})/2i$, we conclude that Eq.~(\ref{eq:15}) determines the stationary current in the channel as $j=j_0 B$, where $j_0=Jd/\hbar$, with $d$ being the lattice period which we set to unity from now on. As expected, $j$ it is proportional to the population difference $\bar{n}_1-\bar{n}_L$ between the two reservoirs and to the relaxation rate $\gamma$, provided that $\gamma \ll J$. It follows from Eq.~(\ref{eq:14}) that the mean number of carries in the lead is $\langle N \rangle = (\bar{n}_1+\bar{n}_L)L/2$. 
This information may be relevant for a numerical simulation of the system dynamics using approximate, e.g., Hilbert-space truncation methods.

Next we analyze the case of a non-vanishing scattering potential. As an example, we consider the potential $V_1$ corresponding to a point-like scatter $E_l=F\delta_{l,L/2}$. Fig.~\ref{fig:3}(c) presents the stationary density matrix for $F=0.5J$. As expected, the diagonal elements $\bar{\sigma}_{l,l}$ now show a jump at the scatterer's position.  Yet, the nearest elements to the main diagonal, which determine the current in the system, have constant value, which is a consequence of the continuity equation. 
We notice non-negligible anti-diagonal matrix elements, which reflect strong spatial correlations between the transmitted and reflected particles.
The analytical expressions for the SPDM from Fig. \ref{fig:3} (c), which is now uniquely characterized by eight quantities, are given in appendix \ref{app:B}.

For a more complicated scattering potential, for example the potential $V_2$ defined by $E_l=F(\delta_{l,L/2-1} + \delta_{l,L/2+1})$, where one meets the phenomenon of resonant scattering, the stationary density matrix has an even more involved structure which is hard to reproduce by means of an algebraic approach. However, we can fairly well reproduce this structure by employing the LB approach \cite{Landau-B} in the next section. 

\subsection{Landauer-B\"uttiker approach}

It is instructive to discuss the result depicted in Fig.~\ref{fig:3}(a) in terms of the Bloch waves $|\kappa \rangle \sim  \sum_l \exp(i\kappa l) |l\rangle$, where we formally consider the limit $L\rightarrow\infty$  and, hence, the quasimomentum $\kappa$ is a continuous quantity.  This limit eliminates the boundary effects and the stationary matrix ${\bar \sigma}_{l,m}$ is approximated by the average matrix $\bar{ \hat \sigma}$,
\begin{eqnarray}
\label{eq:16}
\bar{\hat \sigma}=\int_{-\pi}^{\pi}  | \kappa \rangle\langle \kappa | f(\kappa) \frac{{\rm d} \kappa}{2\pi} \;,
\end{eqnarray} 
where  
\begin{eqnarray}
\label{eq:17}
f(\kappa)=A+2B\sin(\kappa) \;.
\end{eqnarray} 
Remarkably,  the case of a non-vanishing scattering  potential is reproduced by substituting in Eq.(\ref{eq:16}) the Bloch wave $|\kappa\rangle$ by the scattered Bloch waves  $|\kappa'\rangle$:
\begin{eqnarray}
\label{eq:18}
\langle l |\kappa'\rangle\sim \left\{
\begin{array}{ccc}
\exp(i\kappa l) + r(\kappa)\exp(-i\kappa l) &,& l<L/2 \\
t(\kappa)\exp(i\kappa l) &,& l>L/2 
\end{array}  \right. \;.
\end{eqnarray}  
Fig.~\ref{fig:3} compares the exact steady-state SPDM (left panels) with those obtained from Eq.~\eqref{eq:16} (right panels) for the scattering potentials $V_1$ with $F = 0$ (upper panels) and $F = J/2$ (bottom panels). 
It is seen that Eqs.~\eqref{eq:16}-\eqref{eq:18} well approximate the exact SPDM. Weak deviations from the exact result are mainly due to the fact that the LB approach (which is essentially a scattering theory) neglects the dehoherence induced by the (non-perfectly absorbing) reservoirs.

The observed agreement suggests the following equation for the stationary current,
\begin{eqnarray}
\label{eq:19}
j=j_0 \int_{-\pi}^{\pi} \sin(\kappa) f(\kappa) |t(\kappa)|^2 \frac{{\rm d} \kappa}{2\pi} \;, 
\end{eqnarray} 
which follows from Eqs. \eqref{eq:16}-\eqref{eq:18}. In principle, the integration over quasimomentum in Eq.~(\ref{eq:19}) can be substituted by an integration over the energy with ${\rm d}E=J\sin(\kappa){\rm d}\kappa$. Then Eq.~(\ref{eq:19}) takes the same form as the celebrated LB Eq.~(\ref{eq:1}) used to describe particle \cite{Landau-B} and heat transport \cite{Dhar}.  The usage of quasimomentum, however, has the advantage that we obtain not only the current but also reproduce the SPDM.  
\begin{figure}
\includegraphics[width=\linewidth,clip]{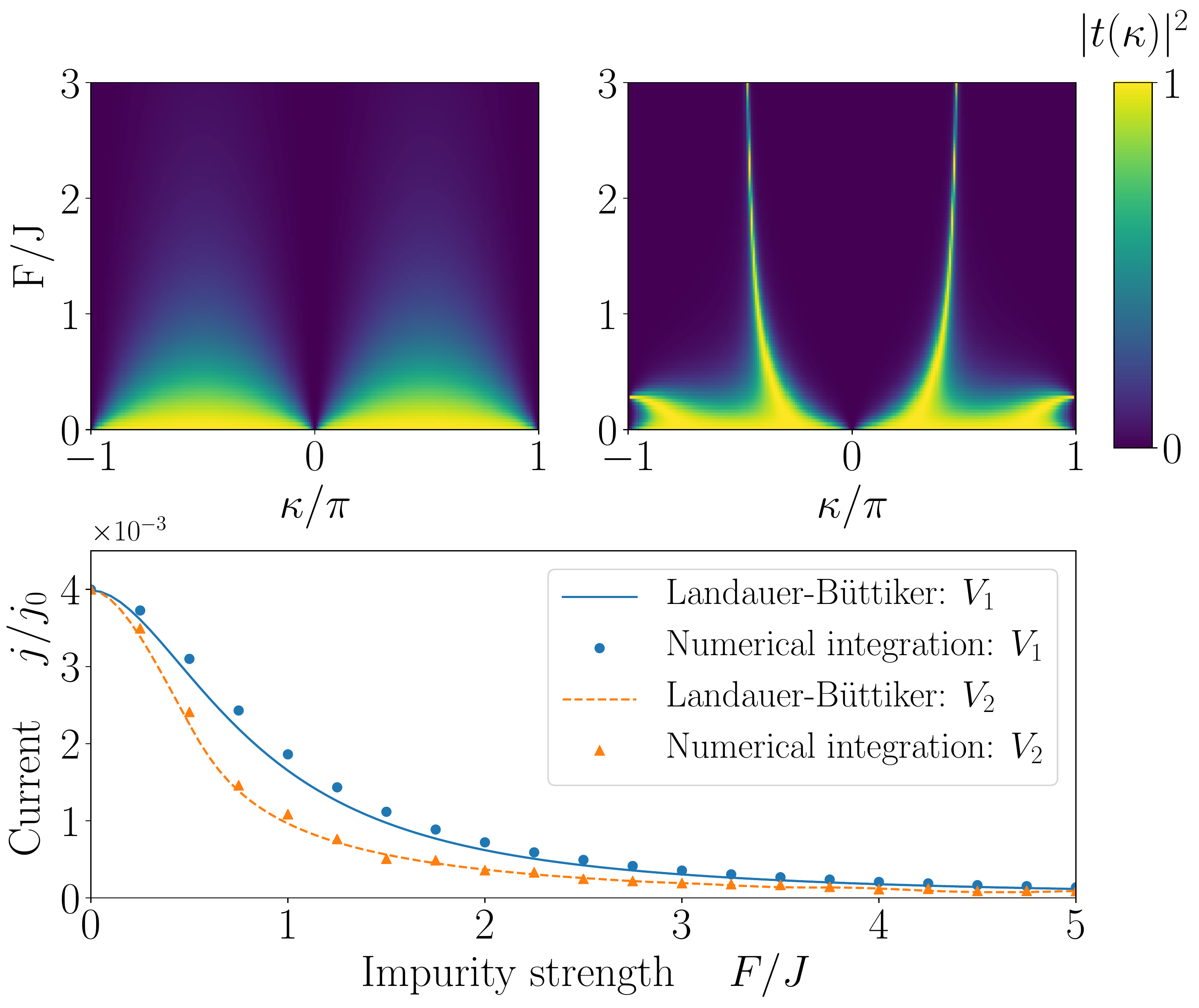} 
\caption{Squared modulus of the transmission coefficient corresponding to the scattering potentials $V_1$ (upper left) and $V_2$ (upper right) as a function of $F/J$ and the quasimomentum $\kappa$. (Bottom panel) Steady state currents across a $L=20$ chain for both potentials and $\bar{n}_1 = 0.02$, $\bar{n}_L = 0$, $\gamma/J = 2$. Results predicted by Eq.~\eqref{eq:19} (solid lines) are compared to exact numerical simulations (symbols).
}
\label{fig:4}
\end{figure} 
The lower panel of Fig.~\ref{fig:4} shows the stationary current across the chain as a function of $F/J$ obtained from the analytical expression \eqref{eq:19} (solid lines), as compared to numerical results based on simulations of the reduced density matrix (symbols) for both above-defined potentials $V_1$ and $V_2$. The squared modulus of the associated transmission coefficients are plotted respectively in the top left and top right panels, as a function of the impurity energy shift $F$ for any quasimomentum $\kappa$. The agreement of the result based on the LB approach with the exact data is very good. 


\section{Conclusions}
We studied the  stationary current of Bose particles between two reservoirs connected by a one-dimensional transport channel. We analyzed the problem from first principles, i.e., without using uncontrolled assumptions. The obtained equation for the stationary current has the same structure as the fermionic LB equation and involves an integration over the energy/quasimomentum of the transmission probability for the carriers weighted by some function $f(\kappa)$.  Similar to the fermionic case, this weight function is determined by the chemical potentials of the reservoirs. Additionally, $f(\kappa)$ was found  to depend on the relaxation constant $\gamma$, whose value is determined by the coupling constant $\varepsilon$ as $\gamma\sim \varepsilon^2$. This result goes beyond the common LB approach because it explicitly takes into account a particular form of coupling between the lead and reservoirs. It should also be mentioned that the obtained Eq.~(\ref{eq:19}) requires infinite temperature of the reservoirs -- the assumption we need to justify the master Eq.~(\ref{eq:5}).  It is an open problem to obtain  the bosonic LB equation for a finite temperature where the Markovian approximation may be not justified and, hence, the equation on the reduced density matrix may include non-Markovian terms. 


\acknowledgements
The authors thank very much Giulio Amato for a critical reading of the manuscript.

\appendix
\section{Transient dynamics of an incoherently pumped double-well}
\label{app1}

\begin{figure}[t]
	\centering
%
%
%
%
%
%
%
%
%
%
%
%
	\includegraphics[width=0.85\linewidth]{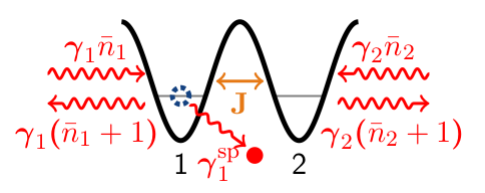}
	\caption{Schematic representation of the open noninteracting double-well.}
	\label{fig:1supp}
\end{figure}

For $U = 0$ and Lindblad operators such as $\hat{L}_\ell=\hat{a}_\ell$ (loss), $\hat{L}_\ell=\hat{a}_\ell^\dagger$ (gain) or $\hat{L}_\ell=\hat{n}_\ell$ (phase noise), the single-particle density matrix's (SPDM) equation of motion \cite{Kordas2015}, Eq.~\eqref{eq:7}, is linear and can thus always be solved by diagonalization. By building a vector $\vec{\sigma}$ out of the SPDM's elements, the equation of motion can be rewritten in the form
\begin{equation}
\partial_t\vec{\sigma}(t) = \mathbf{A}\vec{\sigma}(t)+\vec{b} \,,
\label{eq:A1}
\end{equation}
with $\mathbf{A}$ a diagonally dominant complex symmetric matrix. The general solution of such a system is
\begin{equation}
\vec{\sigma}(t) = e^{\mathbf{A}t}\vec{\sigma}(0)+{\mathbf{A}}^{-1}(\mathbf{1}-e^{\mathbf{A} t})\vec{b} \,,
\label{eq:A2}
\end{equation}
which can be explicitly obtained for a few-site chain. For a time-dependent source term $\vec{b}(t)$, the last term is to be changed into the convolution $\int_{0}^{t}\mathrm{d}t^\prime e^{\mathbf{\mathbf{A}}(t-t^\prime)}\vec{b}(t^\prime)$, which incidentally becomes in the steady state limit a linear superposition of the Laplace transforms of the time-dependent sources. Therefore, everything amounts to calculating \smash{$\mathbf{U}(t) \defeq \exp(\mathbf{A}t) = \mathbf{T}e^{\mathbf{\Lambda} t}\mathbf{T}^{-1}$}.


Let us consider the incoherently pumped double-well depicted in Fig.~(\ref{fig:1supp}). It is described by the following Lindblad master equation \cite{Kordas2015}:
\begin{align}
\partial_t\hat{\rho} = &-i[\hat{H},\hat{\rho}]+\sum_{\mathclap{j=1,2}}\gamma_j^{\mathrm{sp}}\mathcal{D}[\hat{a}_j]\hat{\rho}\nonumber\\ &+\sum_{\mathclap{j=1,2}}\gamma_j\big(\bar{n}_j\mathcal{D}[\hat{a}_j^\dagger]\hat{\rho}
+ (\bar{n}_j+1)\mathcal{D}[\hat{a}_j^{\mathstrut}]\hat{\rho}\big)\label{eq:A3}\\
\hat{H} = &-\frac{J}{2}( \hat{a}^\dagger_{1}\hat{a}_{2} + \hat{a}^\dagger_{2}\hat{a}_{1})
\label{eq:A4}
\end{align}
where $J$ denotes the hopping amplitude, $\gamma_j$ is the incoherent gain-loss process rate at site $j$, and $\gamma_j^\mathrm{sp}$ accounts for a possible supplementary single-particle loss rate at site $j$, modelling engineered loss locally at a site \cite{Ott}, for instance. 
The dissipator superoperator is defined as $\mathcal{D}[\hat{L}]\hat{\rho} = \hat{L}\hat{\rho}\hat{L}^\dagger - 1/2\lbrace\hat{L}^\dagger\hat{L},\hat{\rho} \rbrace$.

By defining $\vec{\sigma}^\mathrm{t} \defeq (\sigma_{11},\sigma_{12},\sigma_{21},\sigma_{22})$, $\Gamma_\ell \defeq 2(\gamma_\ell+\gamma^{\mathrm{sp}}_\ell)/J$, $\bar{\Gamma} \defeq (\Gamma_1+\Gamma_2)/2$ and $\Delta\Gamma \defeq (\Gamma_2-\Gamma_1)/2$, one gets:
\begin{align}
&\mathbf{A}=J\begin{pmatrix}
-\Gamma_1 & i/2 & -i/2 & 0 \\ 
i/2 & -\bar{\Gamma} & 0 & -i/2 \\ 
-i/2 & 0 & -\bar{\Gamma} & +i/2 \\ 
0 & -i/2 & i/2 & -\Gamma_2
\end{pmatrix},
\;
\vec{b}=\begin{pmatrix}\bar{n}_1\gamma_1 \\ 0 \\ 0 \\ \bar{n}_2\gamma_2 \end{pmatrix},\label{eq:A5}\\
&\textrm{Sp}(\mathbf{A}/J)=\{-\tfrac{\bar{\Gamma}}{2}-y,-\tfrac{\bar{\Gamma}}{2}+y,-\tfrac{\bar{\Gamma}}{2},-\tfrac{\bar{\Gamma}}{2}\}\,,
\label{eq:A6}
\end{align}
where $y \defeq \sqrt{\Delta\Gamma^2/4-1}$. 

$\mathbf{A}$ can be diagonalised as $\mathbf{A}=\mathbf{T}\mathbf{\Lambda}\mathbf{T}^{-1}$, where:
\begin{align}
\mathbf{T}&=\begin{pmatrix}\begin{smallmatrix}
1 & 1 & 1 & 0 \\ 
-\tfrac{i}{2}(y+\Delta\Gamma) & \tfrac{i}{2}(y-\Delta\Gamma) & 0 & 1 \\ 
\tfrac{i}{2}(y+\Delta\Gamma) & -\tfrac{i}{2}(y-\Delta\Gamma) & i\Delta\Gamma & 1 \\ 
\tfrac{\Delta\Gamma(y-\Delta\Gamma)}{2}-1 & 1-\tfrac{\Delta\Gamma(y-\Delta\Gamma)}{2} & 1 & 0
\end{smallmatrix}\end{pmatrix},\label{eq:A7}\\
\mathbf{T^{-1}}&=\tfrac{1}{y^2}\begin{pmatrix}\begin{smallmatrix}
-\tfrac{\Delta\Gamma(y-\Delta\Gamma)}{2}-1 & \tfrac{i}{2}(y-\Delta\Gamma) & -\tfrac{i}{2}(y-\Delta\Gamma) & 1 \\ 
\tfrac{\Delta\Gamma(y+\Delta\Gamma)}{2}-1 & -\tfrac{i}{2}(y+\Delta\Gamma) & \tfrac{i}{2}(y+\Delta\Gamma) & 1 \\ 
-2\phantom{\tfrac{|}{}} & i\Delta\Gamma & -i\Delta\Gamma & -2 \\ 
i\Delta\Gamma & (\Delta\Gamma^2-2) & -2 & i\Delta\Gamma
\end{smallmatrix}\end{pmatrix}
\label{eq:A8}
\end{align}
and $\mathbf{\Lambda} = \diag(\{-\bar{\Gamma}-y,-\bar{\Gamma}+y,-\bar{\Gamma},-\bar{\Gamma}\}) $

Finally, the analytical expression for the symmetric $\mathbf{U}(t)$ reads:
\begin{equation}
\label{eq:A9}
\begingroup
\renewcommand*{\arraystretch}{1.}
\mathbf{U}=\tfrac{e^{-\bar{\Gamma}Jt}}{y^2}\begin{pmatrix}
a_{-} & ib_{+}    	& -ib_{+}   & c \\
. & 4y^2-c         & c     & -ib_{-} \\
. & . & 4y^2-c & ib_{-} \\
.  &         .      &   .    & a_{+}
\end{pmatrix}
\endgroup
\end{equation}
where
\begin{align}
a_{\pm} = &-2+(\Delta\Gamma^2-2)\cosh(Jyt)\nonumber\\
&\pm\Delta\Gamma y\sinh(Jy t)\label{eq:A10}\\
b_{\pm} = &\pm\Delta\Gamma(1-\cosh(Jyt))+y\sinh(Jyt)\label{eq:A11}\\
c = &-2+2\cosh(Jyt)\label{eq:A12}
\end{align}
From this matrix and its time integral, one gets the following expressions for the populations:
\begin{widetext}
\begin{align}
n_1(t)=&\frac{n_\textrm{tot}(0)}{2}e^{-\bar{\Gamma}Jt}\bigg\lbrace\left(\frac{4-\Delta\Gamma^2\cosh(Jyt)}{4-\Delta\Gamma^2}-\Delta\Gamma\frac{\sinh(Jyt)}{y}\right)
+\frac{\Delta n(0)}{n_\textrm{tot}(0)}\left(\cosh(Jyt)-\Delta\Gamma\frac{\sinh(Jyt)}{y}\right)\bigg\rbrace\nonumber\nonumber\\
&+f_1(t;\Gamma_1,\Gamma_2)\cdot\gamma_1\bar{n}_1+f_2(t;\Gamma_1,\Gamma_2)\cdot\gamma_2\bar{n}_2\label{eq:A13}\\
n_{2}(t)=&\frac{n_\textrm{tot}(0)}{2}e^{-\bar{\Gamma}Jt}\bigg\lbrace\left(\frac{4-\Delta\Gamma^2\cosh(Jyt)}{4-\Delta\Gamma^2}-\Delta\Gamma\frac{\sinh(Jyt)}{y}\right)
-\frac{\Delta n(0)}{n_\textrm{tot}(0)}\left(\cosh(Jyt)-\Delta\Gamma\frac{\sinh(Jyt)}{y}\right)\bigg\rbrace\nonumber\nonumber\\
&+f_2(t;\Gamma_2,\Gamma_1)\cdot\gamma_1\bar{n}_1+f_1(t;\Gamma_2,\Gamma_1)\cdot\gamma_2\bar{n}_2\label{eq:A14}\\
j(t) =& \,e^{-\bar{\Gamma}Jt}\left(\frac{\Delta\Gamma(1-\cosh(Jyt))}{y^2}n_\mathrm{tot}(0)+\frac{\sinh(yt)}{y}\Delta n(0)\right)+f_3(t;\Gamma_1,\Gamma_2)\cdot\gamma_1\bar{n}_1+f_3(t;\Gamma_2,\Gamma_1)\cdot\gamma_2\bar{n}_2  \,,
\label{eq:A15}
\end{align}
where

\begin{align}
f_1(t;\Gamma_1,\Gamma_2) &= \frac{\Gamma_2+2/\bar{\Gamma}}{\Gamma_1\Gamma_2+4}+e^{-\bar{\Gamma}Jt}\bigg(\frac{2/\bar{\Gamma}}{y^2}-\frac{\Gamma_2+2\bar{\Gamma}/y^2}{\Gamma_1\Gamma_2+4}\cosh(Jyt)+\frac{\Gamma_2(\Gamma_1-\Gamma_2)+2}{\Gamma_1\Gamma_2+4}\frac{\sinh(Jyt)}{y}\bigg)\label{eq:A16}\\
f_2(t;\Gamma_1,\Gamma_2) &= \frac{2/\bar{\Gamma}}{\Gamma_1\Gamma_2+4}+e^{-\bar{\Gamma}Jt}\bigg(\frac{2/\bar{\Gamma}}{y^2}-\frac{2\bar{\Gamma}/y^2}{\Gamma_1\Gamma_2+4}\cosh(Jyt)-\frac{2}{\Gamma_1\Gamma_2+4}\frac{\sinh(Jyt)}{y}\bigg)\label{eq:A17}\\
f_3(t;\Gamma_1,\Gamma_2) &= \frac{\bar{\Gamma}-\Delta\Gamma}{\bar{\Gamma}(\Gamma_1\Gamma_2+4)}+e^{-\bar{\Gamma}Jt}(-\frac{\Delta\Gamma}{y^2}+\frac{\Delta\Gamma(\bar{\Gamma}-\Delta\Gamma)+4\cosh(Jyt)}{y^2(\Gamma_1\Gamma_2+4)} - \frac{\bar{\Gamma}-\Delta\Gamma}{\Gamma_1\Gamma_2+4}\frac{\sinh(Jyt)}{y}) \,.
\label{eq:A18}
\end{align}
\end{widetext}

In the absence of any additional single-body dissipation, the steady state current is maximized for $\gamma_1=\gamma_2=J$, i.e., the frequency of the oscillations of the closed double-well. In this case, the steady-state current is equal to $J(\bar{n}_1-\bar{n}_2)/2$.

\begin{figure}[tb]
	\includegraphics[width=0.9\linewidth]{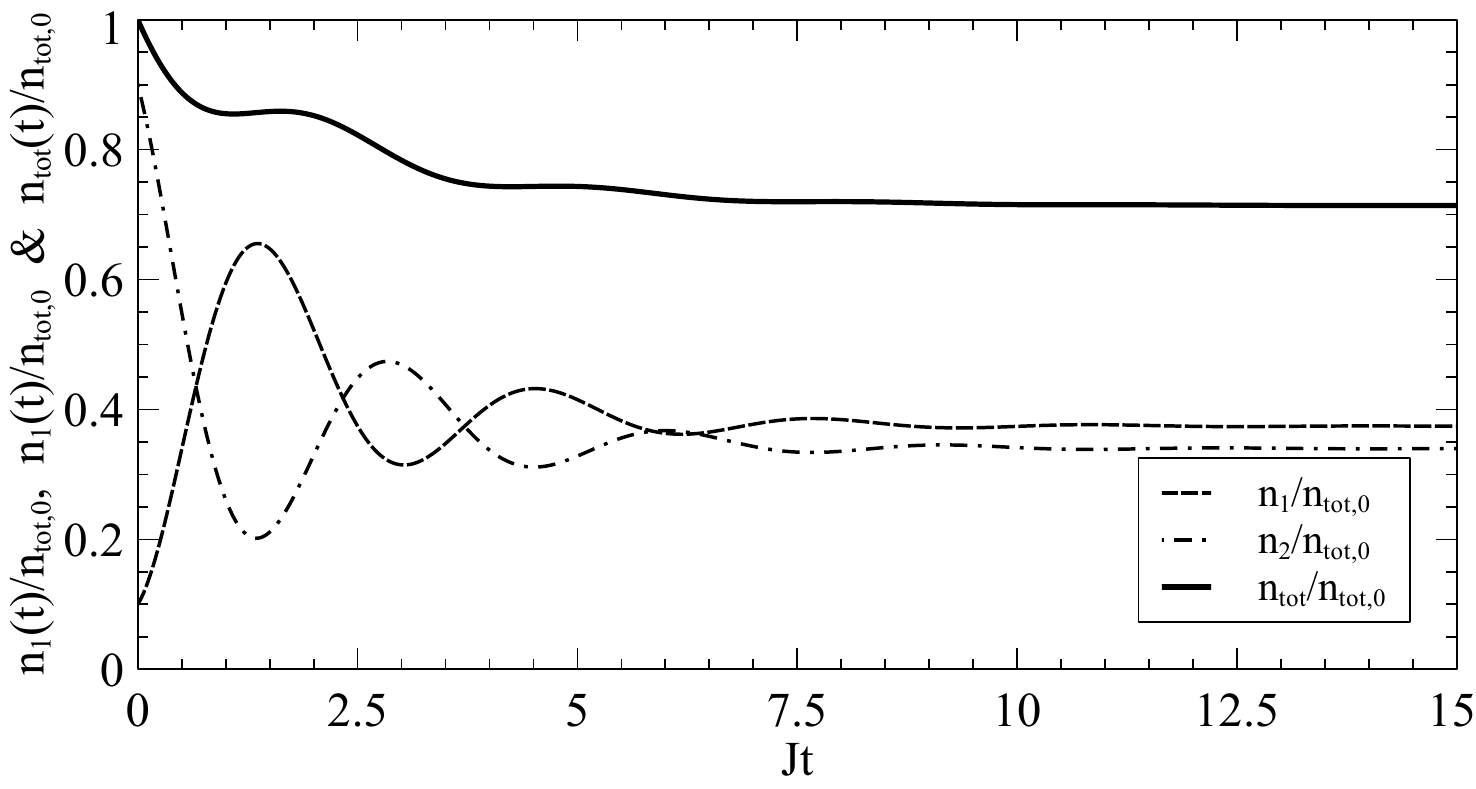}
	\caption{Populations of the double well for an initial imbalance of $\Delta n(0) = -0.8 N_0$, $\gamma_1^\mathrm{sp} = -\gamma_2^\mathrm{sp} = -0.2J$ (the negative rate $\gamma_1^\mathrm{sp}$ is equivalent to the gain dissipator $\gamma_1^\mathrm{sp}\mathcal{D}[\hat{a}_1^\dagger]\hat{\rho}$ for the SPDM), $\gamma_1 = \gamma_2 = 0.5J$, $\bar{n}_1 = 0.5 N_0$, $\bar{n}_2 = 0.2 N_0$.}
	\label{fig:2supp}
\end{figure}

\begin{figure}[H]\centering
		(a)\includegraphics[width=0.9\linewidth]{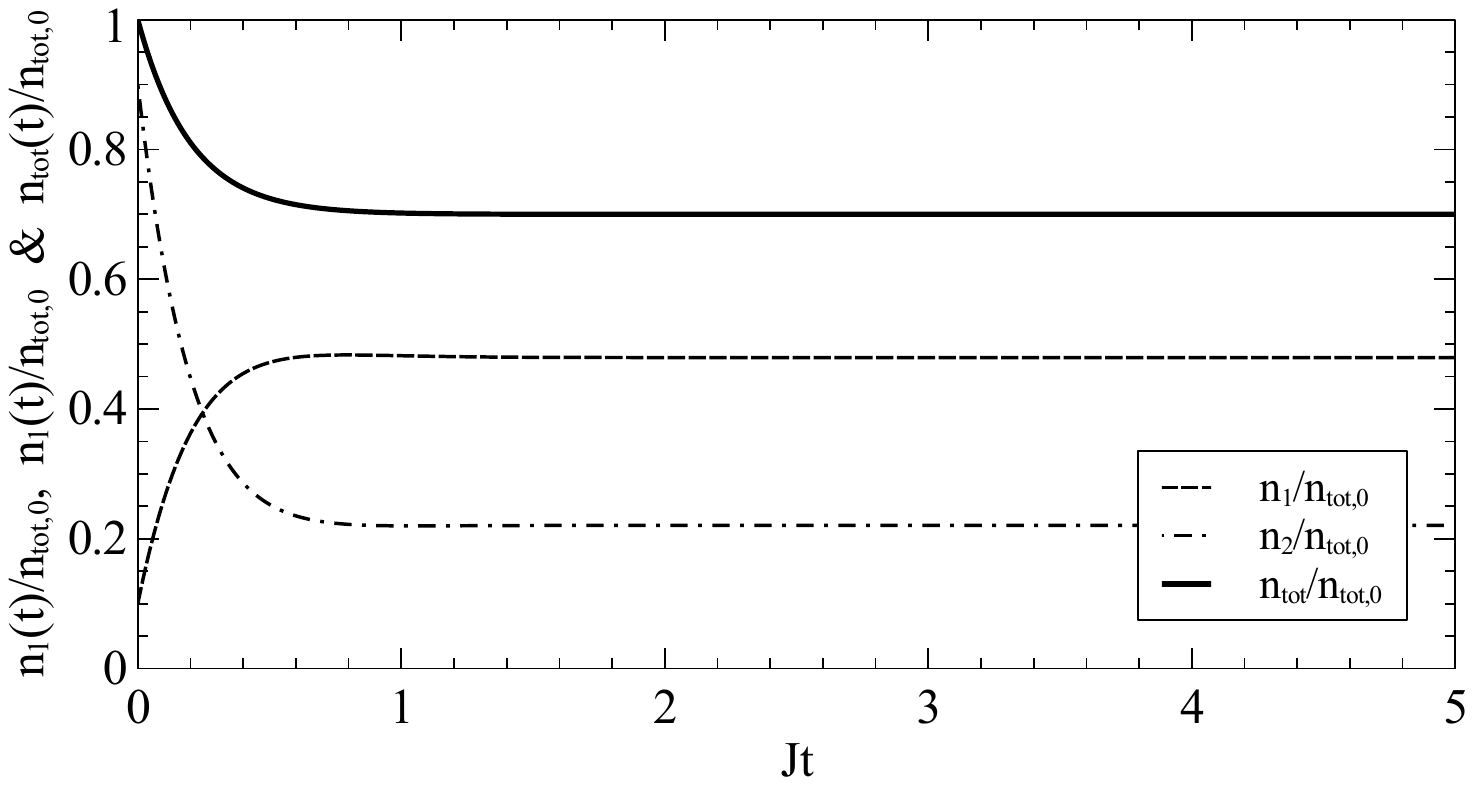} \\
		(b)\includegraphics[width=0.9\linewidth]{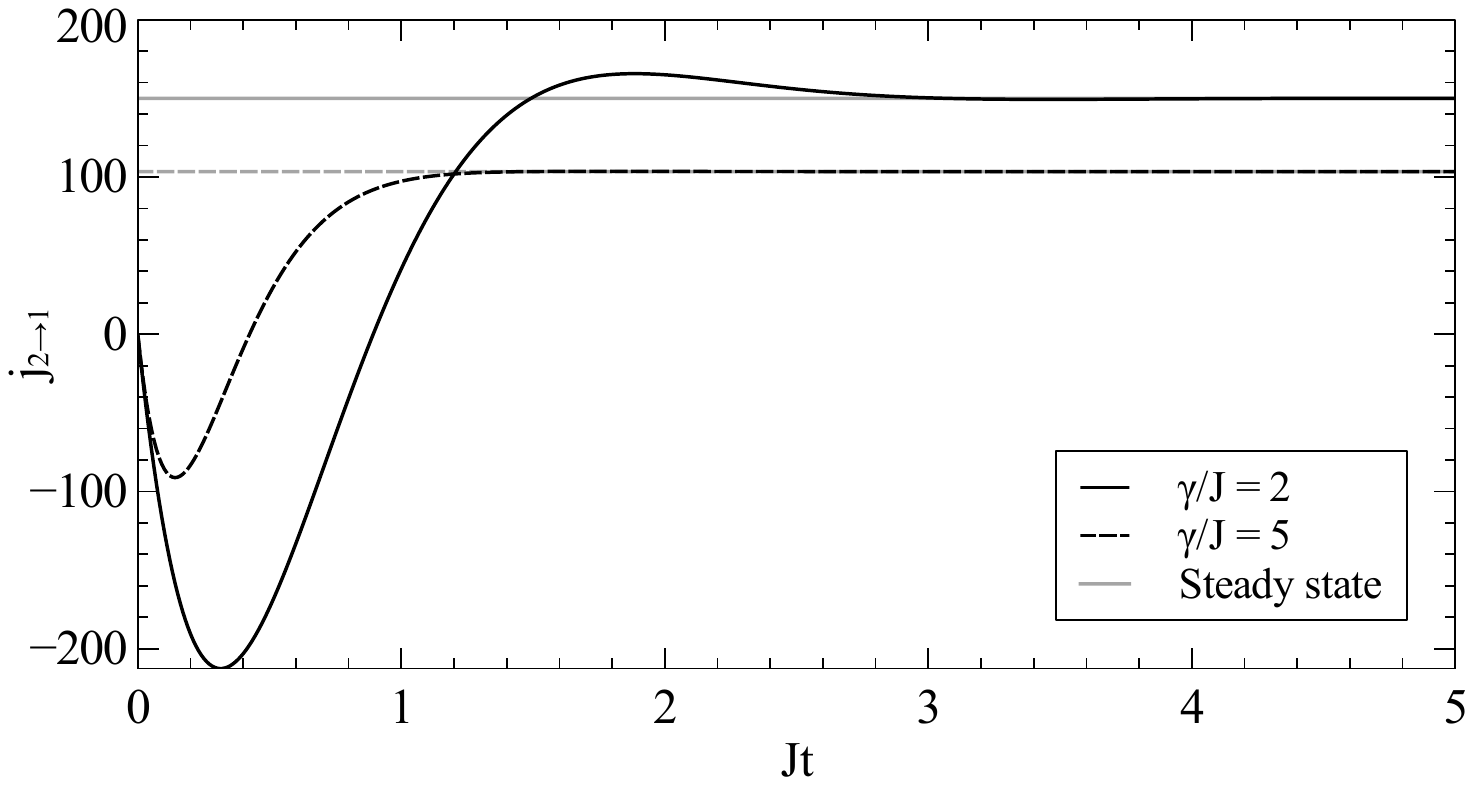}
	\caption{\label{fig:3supp}%
		Double well with an initial imbalance of $\Delta n(0) = -0.8 N_0$. $\gamma_1^{\mathrm{sp}} = \gamma_2^{\mathrm{sp}} = 0$, $\bar{n}_1 = 0.5 N_0$, $\bar{n}_2 = 0.2 N_0$. (a) Populations of the double well for $\gamma_1 = \gamma_2 = 0.5J$. (b) Current from site $2$ to site $1$, $\gamma_1 = \gamma_2 = \gamma$ (see lengend).
	}
\end{figure}

The dynamics derived from the above system of equations is plotted in Figs.~(\ref{fig:2supp}), (\ref{fig:3supp}) (a) and (b).

\section{Analytical solution for the steady-state density matrix in the case of a point-like scatterer}
\label{app:B}

\begin{figure}[t]\vspace{2mm}
\includegraphics[width=\linewidth]{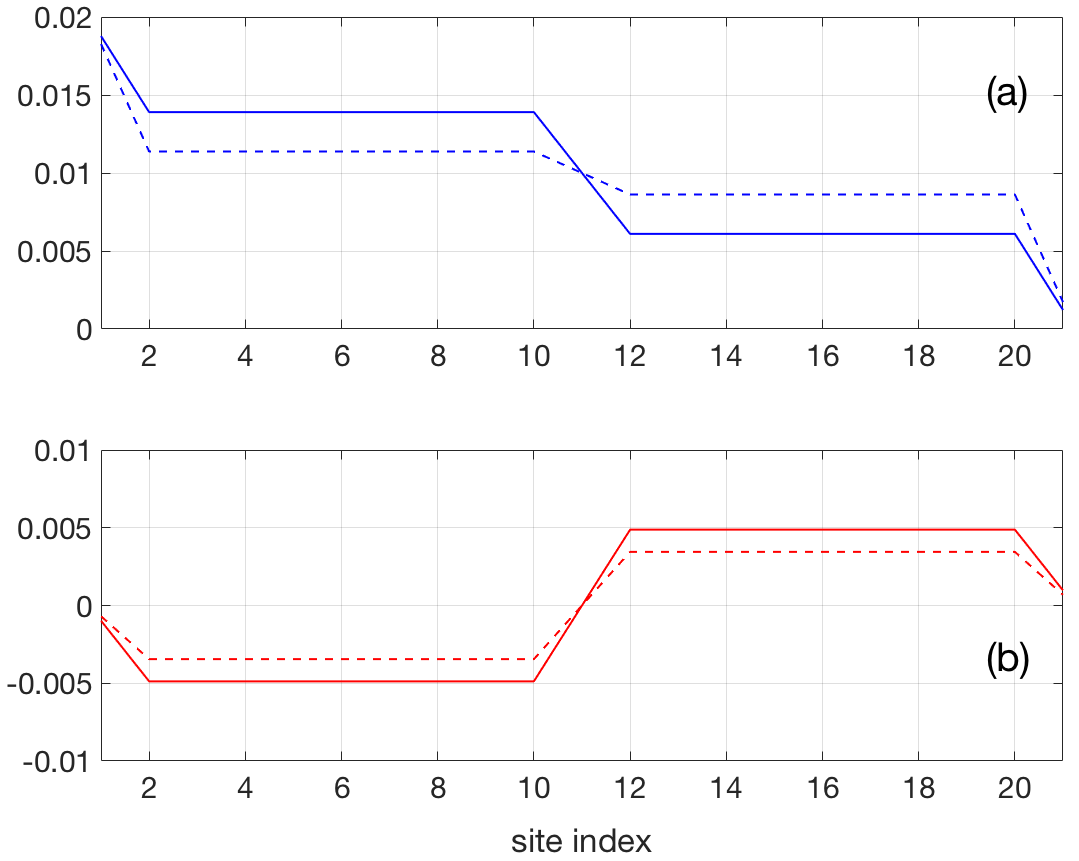} 
\caption{Density matrix elements along the main diagonal (a) and the imaginary part of matrix elements along the main anti-diagonal (b), where the real part is strictly zero except for the central matrix element {$\sigma_{0,0}=(\bar{n}_1+\bar{n}_L)/2$}. Parameters are $L=21$, $\gamma/J=2$, and $F/J=0.5$ (dashed lines) and $F/J=1$ (solid lines). \label{fig:5supp}}
\end{figure}

For the sake of simplicity, we shall consider the case of odd $L$ and shall move the coordinate origin to $l=0$, i.e.  $-M\le l \le M$, where $M \equiv (L-1)/2$. An example of the the steady-state SPDM for $F\ne0$ and $L=21$ is given in Fig.~\ref{fig:3}(c). Formally, this matrix can be viewed as the sum of a tridiagonal matrix and an anti-tridiagonal matrix. The first matrix is characterized by the quantities $C$, $D$, $B$, and $A_1$ and $A_2$,  where the first three quantities have the same meaning as in the case $F=0$, while $A_1$ and $A_2$ denote the occupations of the lattice sites before and after the scatterer, see upper panel of Fig.~\ref{fig:5supp}.  The second matrix takes into account correlations between the transmitted and reflected waves and is hence anti-diagonal. The matrix elements along the main anti-diagonal are pure imaginary and, for the parameters of Fig.~\ref{fig:3}(c), are exemplified in the lower panel in Fig.~\ref{fig:5supp}. We shall characterize them by the quantity $K\equiv i\sigma_{-M,M}=-i\sigma_{M,-M}$ and the quantity $G\equiv -i\sigma_{l,-l}=+i\sigma_{-l,l}$ ($l\ne \pm M$). The first upper and lower anti-diagonals have pure real elements with the values $\pm I$, respectively. Notice that if we sum up these two matrices the four matrix elements around the origin become complex, $\sigma_{\pm 1,0}=\mp I \pm iB$, $\sigma_{0,\pm 1}=\mp I \mp iB$.

Let us now obtain algebraic equations for the introduced quantities. First,  Eqs.~(\ref{eq:10}-\ref{eq:13}) in the main text should be split into two pairs,
\begin{eqnarray}
\label{B6}
\gamma C + JB = \gamma \bar{n}_1 \;,  \\
\gamma B - JC +JA_1 = 0 \;, 
\end{eqnarray} 
and
\begin{eqnarray}
\label{B7}
\gamma B -JA_2 +JD = 0 \;, \\
\gamma D -JB = \gamma \bar{n}_L \;.
\end{eqnarray} 
Similar to the case $F=0$ discussed in the main body of this paper, these equations reflect the matching conditions at the left and right contacts.  Next we match (yet unknown) solutions of the above equations at the scatterer. This gives another three equations,
\begin{eqnarray}
\label{B8}
{\sigma_{0,0} = A_1 + A_2 \;,}\\
{J A_1-J A_2 - 4FI=0 \;,}  \\  
{J G - 2FB=0 \;,}
\end{eqnarray} 
which contain the amplitude of the scattering potential $F$. Finally, we satisfy the matching conditions for the upper-right and lower-left conners of the density matrix, which correspond to the decay of the superposition of the reflected and transmitted waves. We get
\begin{eqnarray}
\label{B9}
\gamma K-JI=0 \;,  \\
JG-JK-\gamma I=0 \;.
\end{eqnarray} 
The obtained full set of eight equations can be easily solved analytically with the following main results: (i) The current through the channel, which is determined by the quantity $B$, is given by
\begin{eqnarray}
\label{B10}
\frac{j}{j_0}\equiv B=\frac{1}{\alpha+ \alpha^{-1} (2F/J)^2}\frac{\bar{n}_1-\bar{n}_L}{2} \;, 
\end{eqnarray} 
where
\begin{eqnarray}
\label{B11}
\alpha=\frac{\gamma}{J}+\frac{J}{\gamma} \;.
\end{eqnarray} 
Thus, asymptotically the current decreases as $(J/F)^2$. This scaling with the energy $F$ of the level of the scattering site implies how the atomic transport may be controlled by locally controlling the energy levels, i.e., by additional electromagnetic potentials in the experiment \cite{Kordas2015, Bloch}. (ii) The population drop at the scatterer is 
\begin{eqnarray}
\label{B12}
{A_1-A_2=8B\alpha^{-1} (F/J)^2\;.}
\end{eqnarray} 
Then, in the limit of large $F$, the population drop approaches the population difference $\bar{n}_1-\bar{n}_L$. 


\end{document}